\documentclass[obeyspaces,preliminary,copyright,creativecommons]{eptcs}
\providecommand{\event}{ICE 2024} % Name of the event you are submitting to

\usepackage{iftex}

\ifpdf
  \usepackage{underscore}         % Only needed if you use pdflatex.
  \usepackage[T1]{fontenc}        % Recommended with pdflatex
\else
  \usepackage{breakurl}           % Not needed if you use pdflatex only.
\fi

\usepackage{localdef}

\usepackage{amsmath}
\usepackage{amsfonts}
\usepackage{amssymb}
\usepackage{latexsym}

\usepackage{graphicx}

\usepackage{listings}
\title{The B2Scala Tool: Integrating Bach in Scala \\with Security in Mind}
\author{Doha Ouardi
\institute{Nadi Research Institute\\
  Faculty of Computer Science\\
  University of Namur \\ Namur, Belgium}
\email{doha.ouardi@unamur.be}
\and
Manel Barkallah 
\institute{Nadi Research Institute\\
  Faculty of Computer Science\\
  University of Namur \\ Namur, Belgium}
\email{manel.barkallah@unamur.be}
\and
Jean-Marie Jacquet 
\institute{Nadi Research Institute\\
  Faculty of Computer Science\\
  University of Namur \\ Namur, Belgium}
\email{\quad jean-marie.jacquet@unamur.be}
}
\def\titlerunning{The B2Scala Tool}
\def\authorrunning{D. Ouardi, M. Barkallah \& J-M. Jacquet}

\hypersetup{
  bookmarksnumbered,
  pdftitle    = {\titlerunning},
  pdfauthor   = {\authorrunning},
  pdfsubject  = {EPTCS},               % Consider adding a more appropriate subject or description
}

\begin{document}
\maketitle

\begin{abstract}
Process algebras have been widely used to verify security protocols in
a formal manner. However they mostly focus on synchronous
communication based on the exchange of messages. We present an
alternative approach relying on asynchronous communication
obtained through information available on a shared space. More
precisely this paper first proposes an embedding in Scala of a
Linda-like language, called Bach.  It consists of a Domain Specific
Language, internal to Scala, that allows us to experiment programs
developed in Bach while benefiting from the Scala eco-system, in
particular from its type system as well as program fragments developed
in Scala. Moreover, we introduce a logic that
allows to restrict the executions of programs to those meeting
logic formulae. Our work is illustrated on the Needham-Schroeder
security protocol, for which we manage to automatically rediscover the
man-in-the-middle attack first put in evidence by G. Lowe.
\end{abstract}

% -------------------------------------------------------------- %

% Introduction
\section{Introduction}

Besides the use of theorem provers, process algebras have been widely
used to verify security protocols in a formal manner. A seminal effort
in this direction is reported in \cite{Lowe}. There the author
illustrates how modeling in CSP~\cite{Hoare-csp} and utilizing the FDR
tool~\cite{FDR} can be used to produce an attack on the
Needham-Schroeder protocol. As another example, the article
\cite{Groote-NDS} demonstrates how state reduction techniques can be
applied to analyze a model of the Bilateral Key Exchange protocol
written in mCRL \cite{mcrl}. In these two cases the models rely on
synchronous communication obtained by the exchange of
messages. Although this type of communication has been fundamental in
the theory of concurrency and has consequently benefited from
extensive research support, it is not necessarily intuitive for
analyzing security protocols. Indeed, the idea of exchanging messages
in a synchronous manner between partners rests on the assumption that
the communication takes place instantaneously on agreed actions and
thus does not naturally leave room for an intruder to intercept
messages. As an evidence at the programming level, in the above two
pieces of work, this has lead the authors to duplicate the exchange of
messages in their model.

Another path of research has been initiated by Gelernter and Carriero,
who advocated in \cite{GC92} that a clear separation between the
interactional and the computational aspects of software components has
to take place in order to build interactive distributed systems. Their
claim has been supported by the design of a model, Linda \cite{CG89},
originally presented as a set of inter-agent communication primitives
which may be added to almost any programming language. Besides process
creation, this set includes primitives for adding, deleting, and
testing the presence/absence of data in a shared dataspace. In doing
so they proposed a new form of synchronization of processes, occurring
asynchronously, through the availability or absence of pieces of
information on a shared space. A number of other models, now referred
to as coordination models, have been proposed afterwards. These models
seem highly attractive to us because, in practice, message exchanges
do not occur atomically through the synchronous communication of
actors. Instead, they must happen through a medium -- such as a
network -- which can be easily modeled as a shared space.

The aim of this paper is to explore how coordination models can
be used to analyze security protocols. More concretely, we will focus
on a specific coordination model, named Bach, will derive a tool,
named B2Scala, and will employ it to produce the attack on the
Needham-Schroeder protocol~\cite{NS} first put in evidence by G. Lowe (see~\cite{Lowe}).

Implementing coordination models can be done in three different
ways. First, as illustrated by Tucson \cite{Tucson}, one may provide
an implementation as a stand alone language. This has the advantage of
offering support for a complete algebra-like incarnation of Linda but
at the expense of having to re-implement classical programming
constructs that are proposed in conventional languages (like
variables, loops, lists, \ldots). The second approach, illustrated by
pSpaces \cite{Loreti-et-al} is to provide a set of APIs in a
conventional language in order to access the shared space through dedicated
functions or methods. This approach benefits from the converse
characteristics of the first one: it is easy to access classical
programming constructs but the abstract control flow that is offered
at a process algebraic level, like non-deterministic choice and
parallel composition, is to be coded in an ad hoc manner.  Finally, a
third approach consists in using a domain specific language embedded
inside an existing language. We will turn to this approach since, in
principle, it enjoys the benefits of the first two approaches.  More specifically, 
this paper proposes to embody the Bach coordination
language inside Scala. In doing so we will profit from the Scala
eco-system while benefiting from all the abstractions offered by the
Bach coordination language. A key feature is that we will interpret
control flow structures, which we put in good use to restrict
computations to those verifying logic formulae. As an interesting consequence, 
we shall be able to produce the man-in-the-middle attack of the 
Needham-Schroeder security protocol first put in evidence by
G. Lowe.

The rest of the paper is structured as
follows. Section~\ref{background} presents the Needham-Schroeder
use-case as well as the Bach and Scala
languages. Section~\ref{btoscala} describes the B2Scala tool, both
from the point of view of its usage by programmers and from the
implementation point of view. A logic is
proposed in Section~\ref{constrainedExecutions} together with its
effect on reducing executions. Section~\ref{solutionNDS} illustrates
how B2Scala coupled to constraint executions can be used to analyze
the Needham-Schroeder protocol. Finally Section~\ref{conclusion} draws
our conclusions and compares our work with related work.

% Background
\section{Background}
\label{background}

\subsection{Use-case: the Needham-Schroeder Protocol}

% \subsection{Use-case : the Needham-Schroeder Protocol}
\label{needham-schroeder-protocol}

The Needham-Schroeder protocol, developed by Roger Needham and Michael
Schroeder in 1978 \cite{NS}, is a pioneering cryptographic solution
aimed at ensuring secure authentication and key distribution within
network environments. Its primary objective is to establish a shared
session key between two parties, typically referred to as the
principal entities, facilitating encrypted communication to safeguard
data confidentiality and integrity. The protocol unfolds in a series
of steps: initialization, where a client (A) requests access to
another client (B) from a trusted server (S), followed by the server's
response, which involves authentication, session key generation, and
ticket encryption. Subsequently, communication with party B ensues,
facilitated by the transmission of the encrypted ticket, along with
nonces to ensure freshness. Parties exchange messages encrypted with
the session key and incorporate nonces to prevent replay
attacks. Mutual authentication is achieved through encrypted messages
exchanged between A and B, leveraging the established session key and
nonces. Despite its early contributions, the original protocol
exhibited vulnerabilities, notably the reflection attack. In response,
refined versions have emerged, such as the
Needham-Schroeder-Lowe \cite{Lowe} and Otway-Rees
protocols \cite{Otway-Rees}.

The description of the Needham-Schroeder public key protocol is often
slimmed down to the three following actions:

\[ \begin{array}{rcll}
   \mathit{Alice} \longrightarrow \mathit{Bob} & : & \mathit{message}(na:a)_{pkb} \\
   \mathit{Bob} \longrightarrow \mathit{Alice}  & : & \mathit{message}(na:nb)_{pka} \\
   \mathit{Alice} \longrightarrow \mathit{Bob} & : & \mathit{message}(nb)_{pkb}
\end{array} \]

\noindent
where each transition of the form $X \rightarrow Y : m$ represents
message $m$ being sent from $X$ to $Y$. Moreover, the notation $m_{k}$
represents message $m$ being encrypted with the public key $k$.

This version assumes that the public keys of Alice and Bob
(resp. $pka$ and $pkb$) are already known to each other.  The full
version also involves communication between the parties and a trusted
server to obtain the public keys.

In this model, Alice initiates the protocol by sending to Bob her
nonce $na$ together with her identity $a$, the whole message being
encrypted with Bob's public key $pkb$.  Bob responds by sending to
Alice her nonce $na$ together with his nonce $nb$, the whole message
being encrypted this time with Alice's public key $pka$. Finally Alice
sends to Bob his nonce $nb$, as a proof that a session has been safely
made between them. The message is this time encrypted with Bob's
public key.

It is worth stressing that, although public keys are known publicly (as
the noun suggests), it is only the owners of the corresponding private
keys that can decrypt encrypted messages. For instance, the first
message sent to Bob can only be decrypted by him.

It is also worth noting that, although sending messages appears as an
atomic action in the above description, this is in fact not the
case. Messages are transmitted through some medium, say 
the network, and thus are subject to be read or picked up by
opponents. This will be illustrated in Section~\ref{solutionNDS} where
a more detailed model will be examined.

\subsection{The Bach Coordination Language}

% \subsection{The Bach Coordination Language}

\Bach~\cite{DJL18,JL07} is a Linda dialect developped at the University of Namur
by the authors. It borrows from Linda the idea of a shared space and
reformulates data and the primitives according to the constraint logic
programming setting~\cite{Saraswat}. The following presentation is based
on the one of article~\cite{BaJa-ICE23}.

\subsubsection{Definition of data}

According to the logic programming setting, we assume a non-empty set
of function names, each one associated with an arity, which indicates
the number of arguments the function takes. We assume a non-empty
subset of function names associated with an arity 0, namely taking no
argument. Such function names are subsequently referred to
as \textit{tokens}. Based on their existence, so-called structured
pieces of information are introduced inductively as expressions of the
form $f(a_1, \cdots, a_n)$ where $f$ is a function name associated
with arity $n$ and where arguments $a_1$, \ldots, $a_n$ are structured pieces of
information, understood either as tokens or in the structured form
under description. Note that, as the special case where $n=0$, tokens
are considered as being structures information terms. The set of
structured pieces of information is subsequently denoted by ${\cal
I}$. For short, \textit{si-term} is used later to denote a structured
piece of information.

\begin{example}
The nounces used by Alice and Bob in the Needham-Schroeder protocol are
coded by the tokens \texttt{na} and \texttt{nb}, respectively. Similarly,
their public keys are coded by the tokens \texttt{pka} and \texttt{pkb}.
A message encrypted by Alice with Bob's public key and providing Alice's nounce
with her identity `$a$' is encoded as the following structured piece of information
\texttt{encrypt(na, a, pkb)}.
\end{example}

% \subsubsection{Agents.}
\subsubsection{Agents}

\begin{figure}[t]
\begin{center}
%\fbox{%
%\begin{minipage}{11cm}
% \begin{scriptsize}
%
\[ \begin{array}{ccc}
    \begin{array}{r@{\hspace*{0.25cm}}c}
     {\bf(T)}
& 
    $$\langle \; tell(t) \; \vert \; \sigma \;\rangle \longrightarrow \langle \; E \;\vert \; \sigma \cup \lbrace t\rbrace \;\rangle$$\\ 
\\ \\ 
 
     {\bf(A)}        
& 
    $$\langle \; ask(t) \;\vert\; \sigma \cup \lbrace t \rbrace \;\rangle \longrightarrow \langle\; E \;\vert\; \sigma \cup \lbrace t\rbrace \;\rangle$$\\ 
% \\ \\ 
% %% 
%      {\bf(G)}        
% & 
%       $$\langle\; get(t)\;\vert\; \sigma \cup \lbrace t \rbrace \;\rangle \longrightarrow \langle\; E \;\vert\; \sigma \;\rangle$$\\
%     \end{array} & \hspace*{1cm} & 
%     \begin{array}{r@{\hspace*{0.25cm}}c}
% % \\ \\ 
%% 
%% 
    \end{array} & \hspace*{1cm} & 
    \begin{array}{r@{\hspace*{0.25cm}}c}
     {\bf(G)}        
& 
      $$\langle\; get(t)\;\vert\; \sigma \cup \lbrace t \rbrace \;\rangle \longrightarrow \langle\; E \;\vert\; \sigma \;\rangle$$\\
\\ \\ 
      {\bf(N)}        
& 
      $$\dfrac{t \not\in \sigma}{\langle\; nask(t)\;\vert\; \sigma \;\rangle \longrightarrow \langle\; E \;\vert\; \sigma \;\rangle}$$\\
    \end{array}
\end{array}
\] 
% \end{scriptsize}
% \end{minipage}
%}
\end{center}

\caption{Transition rules for the primitives (taken from \cite{BaJa-ICE23})}
\label{jmj-fig-primitives}
\end{figure}

Following the concurrent constraint setting, Linda
primitives \texttt{out}, \texttt{rd} and \texttt{in} respectively used
to output a tuple, check its presence and consume one occurrence are
reformulated as \texttt{tell}, \texttt{ask}, \texttt{get}, acting on
si-terms. We add to them a negative counterpart, \texttt{nask}
checking the absence of a si-term. The execution of these primitives
is described by the transition relation defined in
Figure~\ref{jmj-fig-primitives}.  The configurations to be considered
are pairs of instructions, for the moment reduced to simple
primitives, coupled to the contents of the shared space. Following the
concurrent constraint setting, the shared space is referred to as
the \textit{store}. It is taken as a multiset of si-terms. Moreover,
the $E$ symbol is used to denote a terminated computation.
Consequently, rule (T) expresses that the execution of the $tell(t)$
primitive always succeeds and add an occurrence of $t$ to the store.
Rule (A) requires $ask(t)$ to succeed that $t$ is present on the
store. As this primitive just makes a test, the contents of the store
is unchanged. According to rule (G), the $get(t)$ primitive acts
similarly but remove one occurrence of $t$. Finally, as specified by
rule (N), the primitive $nask(t)$ succeeds in case $t$ is absent from
the store.

Primitives are combined to form more complex agents by means of
traditional operators from concurrency theory: sequential composition,
denoted by the $\seqc$ symbol, parallel composition, denoted by the
$\parac$ symbol, and non-deterministic choice, denoted by the
$\choice$ symbol.

Procedures are defined by associating an agent with a procedure name
possibly coupled to parameters. As usual, we shall assume that the
associated agents are guarded, in the sense that the execution of a
primitive preceeds any call or can be rewritten in such a
form. Procedures are subsequently declared after the
\texttt{proc} keyword.

\begin{figure}[t]
% \begin{minipage}{11cm}
% \begin{scriptsize}
\[ \begin{array}{ccc}
    \begin{array}{r@{\hspace*{0.25cm}}c}
     {\bf(S) }
   & 
     \infrulemath{  \transm{ \conf{ A }{ \sigma } 
                      }{ \conf{ A' }{ \sigma' } 
                       } 
            }{ 
                \transm{ \conf{ \seqcc{A}{B} }{ \sigma } 
                      }{ \conf{ \seqcc{A'}{B} }{ \sigma' } 
                       } 
              } 
   \\ \\ 
     {\bf(P) }
   & 
     \infrulemath{  \transm{ \conf{ A }{ \sigma } 
                      }{ \conf{ A' }{ \sigma' } 
                       } 
            }{ 
                 \begin{array}{c} 
                     \transm{ \conf{ \paracc{A}{B} }{ \sigma } 
                           }{ \conf{ \paracc{A'}{B} }{ \sigma' } 
                            } 
                  \\ 
                     \transm{ \conf{ \paracc{B}{A} }{ \sigma } 
                           }{ \conf{ \paracc{B}{A'} }{ \sigma' } 
                            } 
                  \end{array} 
              } 
%     \\ \\ 
%      {\bf(C) }
%    & 
%      \infrulemath{  \transm{ \conf{ A }{ \sigma } 
%                        }{ \conf{ A' }{ \sigma' } 
%                         } 
%             }{ 
%                  \begin{array}{c} 
%                      \transm{ \conf{ \choicec{A}{B} }{ \sigma } 
%                            }{ \conf{ A' }{ \sigma' } 
%                             } 
%                   \\ 
%                      \transm{ \conf{ \choicec{B}{A} }{ \sigma } 
%                            }{ \conf{ A' }{ \sigma' } 
%                             } 
%                   \end{array} 
%               } 
\end{array} & \hspace*{1cm} &     \begin{array}{r@{\hspace*{0.25cm}}c}
% %     \\ \\ 
%      {\bf(Co) }
%    & 
%      \infrulemath{  \models C, \, \transm{ \conf{ A }{ \sigma } 
%                        }{ \conf{ A' }{ \sigma' } 
%                         } 
%             }{ 
%                  \begin{array}{c} 
%                      \transm{ \conf{ C \rightarrow A \diamond B }{ \sigma } 
%                            }{ \conf{ A' }{ \sigma' } 
%                             } 
%                   \\ 
%                      \transm{ \conf{ !C \rightarrow B \diamond A }{ \sigma } 
%                            }{ \conf{ A' }{ \sigma' } 
%                             } 
%                   \end{array} 
%               } 
     {\bf(C) }
   & 
     \infrulemath{  \transm{ \conf{ A }{ \sigma } 
                       }{ \conf{ A' }{ \sigma' } 
                        } 
            }{ 
                 \begin{array}{c} 
                     \transm{ \conf{ \choicec{A}{B} }{ \sigma } 
                           }{ \conf{ A' }{ \sigma' } 
                            } 
                  \\ 
                     \transm{ \conf{ \choicec{B}{A} }{ \sigma } 
                           }{ \conf{ A' }{ \sigma' } 
                            } 
                  \end{array} 
              } 
    \\ \\ 
     {\bf(Pc) }
   & 
     \infrulemath{  P(\overline{x}) = A, \transm{ \conf{ A[\overline{x}/\overline{u} }{ \sigma } 
                       }{ \conf{ A' }{ \sigma' } 
                        } 
            }{ 
                     \transm{ \conf{ P(\overline{u}) }{ \sigma } 
                           }{ \conf{ A' }{ \sigma' } 
                            } 
            }
       \end{array}
\end{array} \]
% \end{scriptsize}
% \end{minipage}

\caption{Transition rules for the operators (taken from \cite{BaJa-ICE23})
\label{fig-operators}}
\end{figure}

The execution of complex agents is defined by the transition rules of
Figure~\ref{fig-operators}.  Sequential, parallel and choice
composition operators are given the convention semantics in rules (S),
(P) and (C), respectively.  Rule (Pc) dictates that the procedure call
$P(\overline{u})$ operates as the agent $A$ that defines $P$
with the formal arguments $\overline{x}$ replaced by the actual ones
$\overline{u}$. It is important to note that, in these rules, agents
of the form ($E ; A$), ($E \parac A$) and ($A \parac E$) are rewritten
as $A$.

\begin{example}
\label{first-coding-Alice-Bob}
As an example, the behavior of Alice and Bob can be coded as follows:

{\normalfont
\begin{lstlisting}
proc Alice = tell(encrypt(na,a,pkb)); get(encrypt(na,nb,pka));
             tell(encrypt(nb,pkb)).

       Bob = get(encrypt(na,a,pkb); tell(encrypt(na,nb,pka));
             get(encrypt(nb,pkb)).
\end{lstlisting}}

\noindent
Note that Alice and Bob only tell messages encrypted with the
public key of the other and only get messages encrypted with their public key,
which simulates their sole use of their private key.

It is also worth stressing that we will present a model of the
Needham-Schroeder protocol and not a concrete implementation. Hence
the above tokens ($na$, $nb$, \ldots) are to be understood as globally
defined and not as a form of local variables.
\end{example}

\subsection{The Scala Programming Language}

% \subsection{The Scala Programming Language}

Scala is a statically typed language known for its concise syntax and
seamless fusion of object-oriented and functional programming. 
Variables can be declared as immutable or mutable, as
illustrated by the following code snippet.

\begin{lstlisting}
val immutableVariable: Int = 42
var mutableVariable: String = "Hello, Scala!"
\end{lstlisting}

\noindent
Methods are introduced with the \textit{def} keyword, can be generic
(with type parameters specified in square brackets), can be written
in curried form (with multiple parameter lists) and have a return type
which is specified at the end of the signature. Here is a simple example
for adding two integers.

\begin{lstlisting}
def add(x: Int, y: Int): Int = x + y
\end{lstlisting}

\noindent
Methods are typically included in the definition of objects, classes
and traits, which act as interfaces in Java. Of particular interest
for the implementation of B2Scala is the definition of \textit{case
  classes} which are classes that automatically define setter, getter,
hash and equal methods.

Two main additional features of Scala are worth stressing.

% \subsubsection{Functions and objects.}
\subsubsection{Functions and objects}

\begin{sloppypar}
Functions may be coded by defining objects with an apply function. For
instance, if we define

\begin{lstlisting}
object tell {
    def apply(siterm: SI_Term) = TellAgent(siterm)
}

object Agent {
	def apply(agent: BSC_Agent) = CalledAgent(agent)
}
\end{lstlisting}

\noindent
then the evaluation of

\begin{lstlisting}
val P = Agent { tell(f(1,2)) }
\end{lstlisting}

\noindent
consists first in evaluating $tell$ on the si-term $f(1,2)$, which
results in the structure $TellAgent(f(1,2))$, and then in evaluating
the function $Agent$ on this value, which results in the structure
$CalledAgent(TellAgent(f(1,2)))$. It is that result which is assigned to $P$.
\end{sloppypar}

% \subsubsection{Strictness and lazyness.}
\subsubsection{Strictness and lazyness}

Scala is a strict language that eagerly evaluates expressions. However
there are cases in which it is desirable to postpone the evaluation of
expressions, for instance to handle recursive definitions of
agents. To that end, Scala proposes two basic mechanisms: call-by-name
of arguments of functions and so-called thunks. To understand these
two concepts, let us modify the \texttt{add} function so that it
returns the double of its first argument, regardless of the value of
the second one:

\begin{lstlisting}
def doubleAdd(x: Int, y: => Int) = x + x
\end{lstlisting}

\noindent
The first argument is passed using the call-by-value strategy. It is evaluated whenever the function is
called. In contrast, the second argument is passed using the
call-by-name strategy. Accordingly, it is evaluated when needed and
thus in our example not evaluated at all.
However one step further needs to be made to handle recursive
expressions that we want to evaluate step by step. In that case,
so-called thunks are used. They amount to consider functions requiring
no arguments, as in the following definition

\begin{lstlisting}
def myIf[A](cond: Boolean, onTrue: () => A,
                           onFalse: () => A): A = {
      if (cond) onTrue() else onFalse()
}
\end{lstlisting}

\noindent
Note that the arguments $onTrue$ and $onFalse$ are functions
taking no arguments and leading to expressions rather than simply
expressions.

To conclude this point, it is possible to delay the evaluation of
val-declared expression by using the $lazy$ keyword, such as in

\begin{lstlisting}
lazy val recursiveExpression =  (1+recursiveExpression)*2
\end{lstlisting}

% B2Scala
\section{The B2Scala Tool}
\label{btoscala}

\subsection{Programming interface}

To embed \Bach\ in Scala, two main issues must be tackled: on the one
hand, how is data declared, and, on the other hand, how are agents
declared.

% \subsubsection{Data.}
\subsubsection{Data}

As regards data, the trait $SI\_Term$ is defined to capture
si-terms. Concrete si-terms are then defined as case classes of this
trait. For instance in order to manipulate $f(1,2)$ in one of the
primitives (tell, ask, \ldots) the following declaration has to be made:

\begin{lstlisting}
case class f(x:Int, y: Int) extends SI_Term
\end{lstlisting}

\noindent
Similarly, tokens can be declared as in

\begin{lstlisting}
case class a() extends SI_Term
\end{lstlisting}

\noindent
However that leads to duplicate parentheses everywhere as in $tell(a())$. To
avoid that a $Token$ class has been defined as a case class of
$SI\_Term$. It takes as argument a string so that token $a$ can be
declared as

\begin{lstlisting}
val a = Token(``a'')
\end{lstlisting}

\noindent
Accordingly, $a$ may now be used without parentheses, as in $tell(a)$.

\begin{example}
As examples, the public keys and nonces used in the Needham-Schroeder protocol
are declared as the following tokens:

{\normalfont
\begin{lstlisting}
val pka = Token(``pka'')
val pkb = Token(``pkb'')
val na = Token(``na'')
val nb = Token(``nb'')
\end{lstlisting}}

\noindent
Encrypted messages are coded by the following si-terms:

{\normalfont
\begin{lstlisting}
case class encrypt2(n: SI_Term,k: SI_Term) extends SI_Term  
case class encrypt3(n: SI_Term,x: SI_Term,k: SI_Term) extends SI_Term
\end{lstlisting}}

\noindent
Note that Scala does not allow the same name to be used for different case
classes. We have thus renamed them according to the number of
arguments.
\end{example}

% \subsubsection{Agents.}
\subsubsection{Agents}  

The main idea for programming agents is to employ constructs of the form

\begin{lstlisting}
val P = Agent { (tell(f(1,2))+tell(g(3))) || (tell(a)+tell(b)) }
\end{lstlisting}

\noindent
which encapsulate a Bach agent inside Scala definitions. The $Agent$
object is the main ingredient to do so. It is defined as an object
with an apply method as follows

\begin{lstlisting}
object Agent {
   def apply(agent: BSC_Agent) = CalledAgent(() => agent)
}
\end{lstlisting}

\noindent
It thus consists of a function mapping a $BSC\_Agent$ into the Scala
structure $CalledAgent$ taking a thunk, which consists of a function
taking no argument and returning an agent. As we saw above, this is needed to
treat in a lazy way recursively defined agent.

The $BSC\_Agent$ type is in fact a trait equipped with the methods
needed to parse Bach composed agents. Technically it is defined as follows:

\begin{lstlisting}
trait BSC_Agent { this: BSC_Agent =>
  def *(other: => BSC_Agent) =
           ConcatenationAgent( () => this, other _)
  def ||(other: => BSC_Agent) =
           ParallelAgent( () => this, other _)
  def +(other: => BSC_Agent) =
           ChoiceAgent( () => this, other _)
}
\end{lstlisting}

\noindent
As $;$ is a reserved symbol in Scala, sequential composition is rewritten
with the $*$ symbol.

The definition of the composition symbol $*$, $||$ and $+$ employs
Scala facility to postfix operations. Using the above definitions, a
construct of the form $tell(t) + tell(u)$ is interpreted as the call
of method $+$ to $tell(t)$ with argument $tell(u)$.

It is worth observing that the composition operators take agent
arguments with call-by-name and deliver structures using thunks,
namely functions without arguments to agents.

It will be useful later to generalize choices such that they offer more than two alternatives according to an index ranging over a set, such as in
\( \sum_{x \in L} ag(x)
\)
where $ag(x)$ is an agent parameterized by $x$. This is obtained in B2Scala by the following construct

\begin{lstlisting}
GSum(L, x => ag(x))
\end{lstlisting}

\noindent
where \texttt{L} is a list.

\subsection{Implementation of the Domain Specific Language}

% To be put back in the final version

The implementation of the domain specific language is based on the
same ingredients as those employed in the Scan and Anemone
workbenches~\cite{Scan,Anemone}. They address two main concerns: how
is the store implemented and how are agents interpreted.

% For anonymous submission
%
% The implementation of the domain specific language has to address two
% main concerns: how is the store implemented and how are agents
% interpreted.

% \subsubsection{The store.}
\subsubsection{The store}

The store is implemented as a mutable map in Scala. Initially empty,
it is enriched for each told structured piece of information by an
association of it to a number representing the number of its
occurrences on the store. The implementation of the primitives follows
directly from this intuition. For instance, the execution of a tell
primitive, say \texttt{tell(t)}, consists in checking whether
\texttt{t} is already in the map. If it is then the number of
occurrences associated with it is simply incremented by one. Otherwise
a new association \texttt{(t,1)} is added to the map. Dually, the
execution of \texttt{get(t)} consists in checking whether \texttt{t}
is in the map and, in this case, in decrementing by one the number of
occurrences. In case one of these two conditions is not met then the
get primitive cannot be executed.

% \subsubsection{Interpretation of agents.}
\subsubsection{Interpretation of agents}
\label{run-one-ref}

Agents are interpreted by repeatedly executing transition steps.  This
boils down to the definition of function \texttt{run\_one}, which
assumes given an agent in an internal form, namely as a subtype of
$BSC\_Agent$, and which returns a pair composed of a boolean and an
agent in internal form. The boolean aims at specifying whether a
transition step has taken place. In this case, the associated agent
consists of the agent obtained by the transition step. Otherwise,
failure is reported with the given agent as associated agent.

The function is defined inductively on the structure of its argument,
say \texttt{ag}. If \texttt{ag} is a primitive, then the \texttt{run\_one}
function simply consists in executing the primitive on the store. 
If \texttt{ag} is a sequentially composed agent $ag_i \seqc
ag_{ii}$, then the transition step proceeds by trying to execute the
first subagent $ag_i$. Assume this succeeds and delivers $ag'$ as
resulting agent. Then the agent returned is $ag' \seqc ag_{ii}$ in
case $ag'$ is not empty or more simply $ag_{ii}$ in case $ag'$ is
empty. Of course, the whole computation fails in case $ag_i$ cannot
perform a transition step, namely in case \texttt{run\_one} applied to
$ag_i$ fails.

The case of an agent composed by a parallel or choice operator is
more subtle. Indeed for both cases one should not always favor the
first or second subagent. To avoid that behavior, we use a boolean variable,
randomly assigned to 0 or 1, and depending upon this
value we start by evaluating the first or second subagent. In case of
failure, we then evaluate the other one and if both fails we report a
failure. In case of success for the parallel composition we determine
the resulting agent in a similar way to what we did for the
sequentially composed agent. For a composition by the choice operator
the tried alternative is simply selected.

The computation of a procedure call is performed by computing the
defining agent.

% Constrained executions
\section{Constrained executions}
\label{constrainedExecutions}

The fact that Bach agents are interpreted in the B2Scala tool opens
the door to select computations of interest. This is obtained by
stating logic formulae to be met.

Two main approaches have been used in concurrency theory to describe
properties by means of logic formulae. One approach, exemplified by
Linear Temporal Logic (LTL)~\cite{Pnueli77}, is based on Kripke
structures. In two words, LTL extends classical propositional logic by
introducing temporal operators that allow to describe how properties
evolve over time. For instance, $X\, \Phi$ means that $\Phi$ holds in
the next state while $\Phi\, U \, \Psi$ specifies that $\Phi$ holds until
$\Psi$ holds. Central to this approach are, on the one hand, a
transition relation between states, indicating which states can be
reached from which states, and, on the other hand, a labelling
function that assigns to each state a set of atomic propositions that
are true in that state.

The other approach is based on labelled transition systems. It is
exemplified by the Hennessy-Milner logic (HML)~\cite{HML}. This logic
provides a way to specify properties in terms of actions and
capabilities. The two following modalities are the key concepts of HML:

\begin{itemize}

\item ${<}a{>}\Psi$ means that, by following the labelled transition
  system, it is possible to make a transition by $a$
  such that the resulting process satisfies $\Psi$

\item $[a]\Psi$ means that, whenever $a$ is performed the resulting
   process satisfies $\Psi$.

\end{itemize}

\noindent
However, since they are finite HML formulae can only describe
properties with a finite depth of reasoning. A way to circumvent this
problem is to use a generalisation called the
$\mu$-calculus~\cite{mu-calculus}. It extends HML with fixed-point
operators, such as in $\mu X. (\Phi \vee {<}a{>}X)$ which states that
there is a path where $\Phi$ holds directly or after having repeatedly
taken $a$-transitions.

The logic we use is inspired by these three logics. It is subsequently
presented in two steps by describing so-called basic formulae and the
bsL-calculus.  The effect on computations is then specified. This
yields so-called constrainted computations.

\subsection{Basic formulae}

Similarly to LTL logic, we first specify
formulae that are true on states. Obviously, a key concept in our
coordination setting is whether a si-term is present on the store under
consideration. This is specified by a construct of the form $\mathit{bf}(t)$
which requires that the si-term $t$ is present on the current
store. The formal definition is as follows.

\begin{definition}
For any si-term $t$, the formula $\mathit{bf}(t)$ holds on store $\sigma$ iff $t \in
\sigma$. This is subsequently denoted as $\sigma \models \mathit{bf}(t)$.
Such formulae are subsequently called \textit{bf-formulae}.
\end{definition}

As expected, bf-formulae can be combined with the classical logic
operators. Formulae built in this way are called \textit{basic formulae}.
The formal definition is as follows.

\begin{definition}
Basic formulae are the formulae meeting the following grammar:
\[ b ::= \mathit{bf}(t) \ | \ !b \ | \ b_1 \vee b_2 \ | \  b_1 \wedge b_2
\]
where $\mathit{bf}(t)$ denotes a bf-formula, $b$, $b_1$, $b_2$ denote
basic formulae and the symbols $!$, $\vee$, $\wedge$ respectively express
the negation, the disjunction and the conjuction of basic
formulae.
\end{definition}

The fact that a basic formula $f$ holds on the store $\sigma$ is
defined from the relation $\models$ on bf-formulae according to the
traditional truth tables of propositional logic. By extension, this
will be subsequently denoted by $\sigma \models f$.

\begin{sloppypar}
\begin{example}
As an example, $\mathit{bf}(i\_running(Alice,Bob))$ is a bf-formula that
states that the si-term $i\_running(Alice,Bob)$ is on the store, which
can be used to specify that Alice and Bob have initiated a session.
\end{example}
\end{sloppypar}

\subsection{The bsL calculus}

Similarly to Hennessy-Milner logic and the mu-calculus, we now turn to
specify sequences of properties that have to hold on the sequences of
stores produced by computations. Obviously, as we want to restrict
computations, we have to discard the $[\ldots]$ modality. However we
can use the ${<}\ldots{>}$ modality in the following manner. Remember that
in HML the formula ${<}a{>}{<}b{>}F$ expresses that it is possible to do an
$a$ step followed by a $b$ step and reach a process in which $F$
holds. In a similar way, we will express by $\mathit{bf}(a);\mathit{bf}(b)$ the property
that it is possible to do a step which leads to $\mathit{bf}(a)$ being true
followed by a step after which $\mathit{bf}(b)$ is true. This is for instance
performed by the Bach agent $tell(a); tell(b)$. Note that as a
reminder of the sequential composition of agents in Bach, we have used
the ``;'' to compose sequentially bf-formulae. As noticed in the above
mu-calculus formula, besides sequential composition, we shall also use
disjunction to allow the choice between several paths. This leads us
to the following grammar where, by analogy to Bach operators, the
``+'' symbol is used to indicate disjunction.

\begin{definition}
BsL-formulae are the formula defined by the following grammar:
\begin{eqnarray*}
  f & ::= & b \ | \ P \ | \ f_1 \choice f_2 \ | \ f_1 \seqc f_2
\end{eqnarray*}
where $b$ denotes a basic formula, $f_1$ and $f_2$ are bsL-formulae and
$P$ a variable to be defined by an equation of the form $P =
f'$ with $f'$ being a bsL-formula. As usual in concurrency theory, we assume that $f'$ is guarded in the
sense that a bf-formula is requested before variable $P$ is
called recursively.
\end{definition}

\begin{example}
As an example, the attack on the Needham-Schroeder protocol may be
discovered by finding a computation that obeys the bsL-formula
$X$ defined by
\begin{eqnarray*}
  X & = & (\, not(i\_running(Alice,Bob)) \seqc X \, ) + r\_commit(Alice,Bob) 
\end{eqnarray*}
that is by a computation that does not produce the si-term
$i\_running(Alice,Bob)$ and that ends when $r\_commit(Alice,Bob)$
appears on the store. Restated in other terms such a computation never
includes the start of a session between Alice and Bob but terminates
with Alice and Bob ending the session by committing together.
\end{example}

\subsection{Constrained computations}

We are now in a position to detail how computations may be constrained
by bsL-formulae.  Intuitively, if $f$ is a bsL-formula composed of a
sequence of basic formulae, a computation $c$ is considered to be
constrained by $f$ if the sequence of stores involved in $c$
successively obeys the successive basic formulae in $f$. This is
defined by means of the auxiliary $\vdash$ relation, itself defined by
the rules of Figure~\ref{fig-bHM-deriv}. Intuitively, the notation
\( \bHMderiv{\sigma}{f}{f'} \) states that a first basic formula of $f$ is
satisfied on the store $\sigma$ and that the remaining formulae of
$\mathit{f'}$ need to be satisfied. Accordingly rule (BF) asserts that
if the basic formula $b$ is satisfied by the store $\sigma$ then it is
also the first formula to be satisfied and nothing remains to be
established. The symbol $\epsilon$ is used there to denote an empty
sequence of basic formulae. Rule (PF) states that if formula $P$ is
defined as $\mathit{f}$ and if a first bf-formula of $\mathit{f}$ is
satisfied by $\sigma$ yielding $\mathit{f'}$ to be satisfied next then
so does $P$ with $\mathit{f'}$ to be satisfied next. Finally
rules (CF) and (SF) specify the choice and sequential composition of
bsL-formulae as one may expect.

\begin{figure}[t]
\begin{center}
%\fbox{%
% \begin{minipage}{9cm}
% \begin{scriptsize}
%
\[ \begin{array}{ccc}
    \begin{array}{r@{\hspace*{0.25cm}}c}
     {\bf(BF)}
& 
   $$\dfrac{ \sigma \models b }{ \bHMderiv{\sigma}{b}{\epsilon} }$$\\ 
    \\ \\
     {\bf (CF)}        
& 
     $$\dfrac{ \bHMderiv{\sigma}{f_1}{f_3} }{
       \begin{array}{c}
         \bHMderiv{\sigma}{(f_1 + f_2)}{f_3} \\
         \bHMderiv{\sigma}{(f_2 + f_1)}{f_3} \\         
       \end{array} }$$\\ 
% \\ \\ 
    \end{array} & \hspace*{1cm} & 
    \begin{array}{r@{\hspace*{0.25cm}}c}
     {\bf(PF)}        
& 
      $$\dfrac{ P = f, \hspace*{2ex} \bHMderiv{\sigma}{f}{f'} }{ \bHMderiv{\sigma}{P}{f'} }$$\\
\\ \\ 
     {\bf(SF)}        
& 
     $$\dfrac{ \bHMderiv{\sigma}{f_1}{f_3} }{
         \bHMderiv{\sigma}{(f_1 \seqc f_2)}{(f_3 \seqc f_2)} }$$\\ 
    \end{array}
\end{array}    
\] 
%\end{scriptsize}
% \end{minipage}
%}
\end{center}

\caption{Transition rules for the $\vdash$ relation}
\label{fig-bHM-deriv}
\end{figure}

Given the $\vdash$ relation, we can define constrained computations by
extending the $\rightarrow$ transition relation as the
$\hookrightarrow$ relation specified by rule (ET) of
Figure~\ref{ext-transition}. Informally this rule states that if, on
the one hand, agent $A$ can do a transition from the store $\sigma$
yielding a new agent $A'$ and a new store $\sigma'$ and if, on the
other hand, a first formula of $f$ is met by $\sigma'$ yielding
$f'$ as a remaining bsL-formula to be established, then agent $A$
can make a constrained transition from store $\sigma$ and bHM-formula
$f$ to agent $A'$ to be computed on store $\sigma'$ and with respect
to bHM-formula $f'$.

\begin{figure}[t]
\begin{center}
%\fbox{%
% \begin{minipage}{9cm}
% \begin{scriptsize}
%
\[ \begin{array}{r@{\hspace*{0.25cm}}c}
     {\bf(ET)}
& 
     $$\dfrac{ \conf{A}{\sigma} \longrightarrow \conf{A'}{\sigma'}, \hspace*{2ex}
               \bHMderiv{\sigma'}{f}{f'}
               }{ \extconf{A}{f}{\sigma} \hookrightarrow \extconf{A'}{f'}{\sigma'} }$$
\end{array}
\] 
%\end{scriptsize}
% \end{minipage}
%}
\end{center}

\caption{Extended transition rule}
\label{ext-transition}
\end{figure}

It is worth noting that the encoding in B2Scala is quite easy. On the
one hand, bf-formulae are defined similarly to Bach primitives through the
\texttt{bf} function and are combined as primitives are. On the other
hand, bsL formulae are defined by the \texttt{bsL} function and
recursive definitions are handled in the same way as recursive agents.

The interpretation of agents is then made with respect to a
bsL-formula.  Basically, a step is allowed by the \texttt{run\_one}
function if one step can be made according to the bsL-formula, as
specified by the $\hookrightarrow$ transition relation. This results
in a new agent to be solved together with the continuation of the
bsL-formula to be satisfied.

% Applications
\section{The Needham-Schroeder protocol in B2Scala}
\label{solutionNDS}

% To be reworked for the last version
% 
% As an application of the B2Scala tool, let us now code the
% Needham-Schroeder protocol and exhibit a computation that reflects
% G. Lowe's attack. The interested reader will find the code, the tool
% and a video of its usage under the web pages of the authors at the
% addresses mentioned in \cite{NDS-protocol-in-bscala}.

As an application of the B2Scala tool, let us now code the
Needham-Schroeder protocol and exhibit a computation that reflects
G. Lowe's attack. The interested reader will find the code, the tool
and a video of its usage under the web pages of the authors at the
addresses mentioned in \cite{NDS-protocol-in-bscala}.

Allowing for an attack requires us to introduce an intruder. It is
subsequently named Mallory. This being said, the first point to
address is to declare nonces and public keys for all the participants
of the protocol, namely Alice, Bob and Mallory. This is achieved by
the following token declarations:

\begin{lstlisting}
val na = Token("Alice_nonce")
val nb = Token("Bob_nonce")
val nm = Token("Mallory_nonce")

val pka = Token("Alice_public_key")
val pkb = Token("Bob_public_key")
val pkm = Token("Mallory_public_key")    
\end{lstlisting}

It will also be useful later to refer to the three participants, which
can be achieved by means of the following token declarations:

\begin{lstlisting}
val alice   = Token("Alice_as_agent")
val bob     = Token("Bob_as_agent")
val mallory = Token("Mallory_as_intruder")
\end{lstlisting}

To better view who takes which message produced by whom, encrypted
messages introduced in Section~\ref{btoscala}, are slightly extended
as si-terms of the form
$\mathit{message(Sender,Receiver,Encryted\_Message)}$. Moreover, to
highlight which message is used in the protocol, we shall subsequently
rename encrypted messages as $\mathit{encrypt}\_n$, with $n$ the
number in the sequence of messages. This has the additional advantage of
avoiding to overload case classes, which is forbidden in Scala. The following
declarations follow.

\begin{lstlisting}
case class encrypt_i(vNonce: SI_Term, vAg: SI_Term,
                        vKey: SI_Term) extends SI_Term
case class encrypt_ii(vNonce: SI_Term, wNonce: SI_Term,
                        vKey: SI_Term) extends SI_Term
case class encrypt_iii(vNonce: SI_Term,
                        vKey: SI_Term) extends SI_Term
case class message(agS: SI_Term, agR: SI_Term,
                        encM: SI_Term) extends SI_Term
\end{lstlisting}

Finally, si-terms are introduced to indicate with whom Alice and Bob
start and close their sessions. They are declared as follows:

\begin{lstlisting}
case class a_running(vAg: SI_Term) extends SI_Term
case class b_running(vAg: SI_Term) extends SI_Term
case class a_commit(vAg: SI_Term) extends SI_Term
case class b_commit(vAg: SI_Term) extends SI_Term
\end{lstlisting}

\begin{figure}[t]
\begin{lstlisting}
val Alice = Agent {
  GSum( List(bob,mallory), Y => {
     tell(a_running(Y)) *
     tell( message(alice, Y, encrypt_i(na, alice, public_key(Y))) ) *
     GSum( List(na,nb,nm),  WNonce => {
        get( message(Y, alice, encrypt_ii(na,WNonce,pka)) ) *
        tell( message(alice,Y,encrypt_iii(WNonce,public_key(Y))) ) *
        tell( a_commit(Y) )
     })
  })
}
\end{lstlisting}

\caption{Coding of Alice in B2Scala}
\label{Alice-in-bscala}
\end{figure}

We are now in a position to code the behavior of Alice, Bob and
Mallory. Coding Alice's behavior follows the description we gave in
Example~\ref{first-coding-Alice-Bob} in Section~\ref{background}. The
code is provided in Figure~\ref{Alice-in-bscala}. Although Alice wants
to send a first encrypted message to Bob, she can just put her message
on the network, hoping that it will reach Bob. The network is
simulated here by the store, which leaves room to Mallory to intercept
it. As a result, the first action is for Alice to start of a session.
Hopefully it is with Bob but, to test for a possible attack, we have
to take into account the fact that Mallory can take Bob's place. This
is coded by offering a choice between Bob and Mallory by the
\texttt{GSum([bob,mallory], \ldots)} construct. Calling this actor
$Y$, Alice's first action is to tell the initialization of the session
with $Y$, thanks to the \texttt{a\_running(Y)} si-term being told and
then to tell the first encrypted message with her nonce, her identity
and the public key of Y. The sender and receiver of this message are
respectively $Alice$ and $Y$. Then Alice waits for a second encrypted
message with her nonce and what she hopes to be Bob's nonce, this
message being encrypted by her public key. As the second nonce is
unknown a new choice is offered with the $\mathit{WNonce}$ si-term. Finally,
Alice sends the third encrypted message with this nonce, encoded with
the public key of $Y$ and terminates the session by telling the
\texttt{a\_commit(Y)} si-term. It is worth noting that
\texttt{public\_key(Y)} consists of a call to a Scala function that
returns the public key corresponding to the $Y$ argument.

Coding Bob's behavior proceeds in a dual manner. This time the coding
has to take into account that Mallory can have taken Alice's
place. Hence the first choice \texttt{GSum([alice,mallory], \ldots)}
with $Y$ denoting the sender of the message. Moreover, the identity of
the agent in the first message being got can be different from $Y$. A
second choice \texttt{GSum([alice,mallory], \ldots)} results from
that. The whole agent is given in Figure~\ref{Bob-in-bscala}.

\begin{figure}[t]
\begin{lstlisting}
val Bob = Agent {
   GSum( List(alice,mallory),  Y => { 
      tell(b_running(Y)) *
      GSum( List(alice,mallory), VAg => {
         get( message(Y,bob,encrypt_i(na,VAg,pkb)) ) *
         tell( message(bob,Y,encrypt_ii(na,nb,public_key(VAg))) ) *
         get( message(Y,bob,encrypt_iii(nb,pkb)) ) *
         tell( b_commit(VAg) )
      })
   })
}
\end{lstlisting}

\caption{Coding of Bob in B2Scala}
\label{Bob-in-bscala}
\end{figure}

As an intruder, Mallory gets and tells messages from Alice and Bob,
possibly modifying some parts in case the messages are encrypted with
his public key. This applies for the three kinds of message
sent/received by Alice and Bob. Figure~\ref{Mallory-in-bscala}
provides the code for the first message. It presents three
\texttt{GSum} choices resulting from the three unknown arguments
\texttt{VNonce}, \texttt{VAg}, \texttt{VPK} of the message.  In all
the cases, Bob's attitude is to get the message and to resend it, by
modifying the public key if he can decrypt the message, namely if
\texttt{VPK} is his public key.

\begin{figure}[t]
\begin{lstlisting}
lazy val Mallory:BSC_Agent = Agent {

   ( GSum( List(na,nb,nm), VNonce => {
      GSum( List(alice,bob), VAg => {
       GSum( List(pka,pkb,pkm), VPK => {
         get( message(alice,mallory,encrypt_i(VNonce,VAg,VPK)) ) *
         ( if ( VPK == pkm) {
             tell( message(mallory,bob,encrypt_i(VNonce,VAg,pkb)) )
           } else {
             tell( message(mallory,bob,encrypt_i(VNonce,VAg,VPK)) )		 
	   } ) * Mallory
	})
       })
      }) )     + ...
\end{lstlisting}

\caption{Coding of Mallory in B2Scala}
\label{Mallory-in-bscala}
\end{figure}

To conclude the encoding of the protocol in B2Scala, a bsL-formula $F$
is specified, on the one hand, by excluding a session starting between
Bob and Alice and, on the other hand, by requiring the end of the
session by Bob with Alice. These two requirements are obtained through
the basic formulae \texttt{inproper\_init} and \texttt{end\_session},
as specified below:

\begin{lstlisting}
val inproper_init = not( bf(a_running(bob)) or bf(b_running(alice)) )
val end_session = bf(b_commit(alice))
\end{lstlisting}

Formula $F$ is then coded recursively by requiring $F$ after a step meeting
\texttt{inproper\_init} and by stopping the computation once a step is done that
makes \texttt{end\_session} holds. This is specified as follows.

\begin{lstlisting}
val F = bsL { (inproper_init * F) + end_session }
\end{lstlisting}

Computations are started by invoking the following Scala instructions

\begin{lstlisting}
val Protocol = Agent { Alice || Bob || Mallory }

val bsc_executor = new BSC_Runner
bsc_executor.execute(Protocol,F)
\end{lstlisting}

\begin{figure}[t]
\begin{center}
\includegraphics[width=16cm]{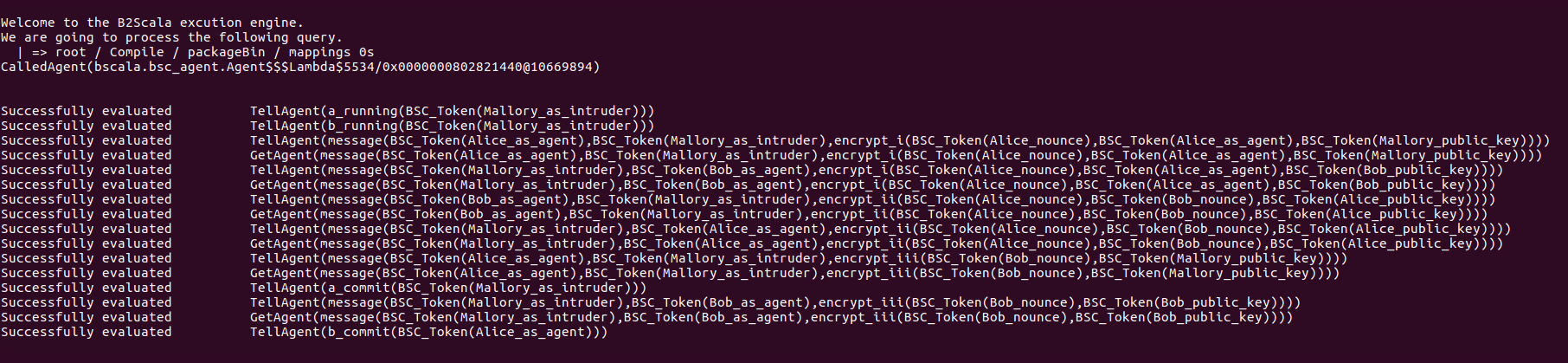}
\caption{Screenshot of the computation
  \label{computation-result}}
\end{center}
\end{figure}

The result is given in Figure~\ref{computation-result} in a verbose
form in which all the primitives are displayed as Scala objects. As we
shall see in a few lines, it produces G. Lowe's attack. To view that,
let us reformulate the Scala objects $TellAgent$, $GetAgent$ and
$BSC\_Token$ in their corresponding Bach counterparts. The listing of 
Figure~\ref{computation-result} then becomes as follows, where numbers are
introduce to facilitate the explanation:

\begin{lstlisting}
 (1)  tell(a_running(mallory))
 (2)  tell(b_running(mallory))
 (3)  tell(message(alice,mallory,encrypt_i(na,alice,pkm)))
 (4)  get(message(alice,mallory,encrypt_i(na,alice,pkm)))
 (5)  tell(message(mallory,bob,encrypt_i(na,alice,pkb)))
 (6)  get(message(mallory,bob,encrypt_i(na,alice,pkb)))
 (7)  tell(message(bob,mallory,encrypt_ii(na,nb,pka)))
 (8)  get(message(bob,mallory,encrypt_ii(na,nb,pka)))
 (9)  tell(message(mallory,alice,encrypt_ii(na,nb,pka)))
(10)  get(message(mallory,alice,encrypt_ii(na,nb,pka)))
(11)  tell(message(alice,mallory,encrypt_iii(nb,pkm)))
(12)  get(message(alice,mallory,encrypt_iii(nb,pkm)))
(13)  tell(a_commit(mallory))
(14)  tell(message(mallory,bob,encrypt_iii(nb,pkb)))
(15)  get(message(mallory,bob,encrypt_iii(nb,pkb)))
(16)  tell(b_commit(alice))
\end{lstlisting} 

Lines (1), (2), (13) and (16) evidence that Alice and Bob have
actually exchanged messages with Mallory whereas they thought they
would speak to each other. In fact Mallory manages to make himself
appear as Bob to Alice and as Alice to Bob. Let us abstract from these
lines. It is then worth observing that the above listing makes appear
tell and get in pairs employing the same message. This
corresponds to one actor sending the message to another actor, which
is translated in our framework as the first actor telling the message and
the second one getting it. By reusing the description of
Section~\ref{needham-schroeder-protocol}, the listing can then be
summed up as follows:

\[ \begin{array}{rcllll}
   \mathit{Alice} \longrightarrow \mathit{Mallory} & : & \mathit{message}(na:alice)_{pkm} & \hspace*{1.5cm} & \mbox{\small (lines 3 and 4)}\\
   \mathit{Mallory} \longrightarrow \mathit{Bob} & : & \mathit{message}(na:alice)_{pkm} & & \mbox{\small (lines 5 and 6)} \\
   \mathit{Bob} \longrightarrow \mathit{Mallory}  & : & \mathit{message}(na:nb)_{pka} & & \mbox{\small (lines 7 and 8)} \\
   \mathit{Mallory} \longrightarrow \mathit{Alice}  & : & \mathit{message}(na:nb)_{pka} & & \mbox{\small (lines 9 and 10)} \\
   \mathit{Alice} \longrightarrow \mathit{Mallory} & : & \mathit{message}(nb)_{pkm} & & \mbox{\small (lines 11 and 12)} \\
   \mathit{Mallory} \longrightarrow \mathit{Bob} & : & \mathit{message}(nb)_{pkb} & & \mbox{\small (lines 14 and 15)} \\
\end{array} \]

\noindent
This is in fact the attack identified by G. Lowe in \cite{Lowe}.
It consists essentially in placing Mallory in between
Alice and Bob, in having him forward Alice's first message, by changing the
public key encrypting the message, in getting Bob's reply and transfer it as
such, and finally in forwarding Alice's reply to Bob, again by changing the
public key encrypting the message.

Note that a key ingredient for producing the above computation is that
imposing \texttt{inproper\_init} to hold forces the first choice in
Alice's code and Bob's code to be made such that $Y$ takes Mallory as
value.

% % Comparison with related work
% \input{related_work.tex}

% Conclusion
\section{Conclusion}
\label{conclusion}

% To be reused for final version
% 
% As a complementary line to previous work \cite{Scan,Anemone}, this
% paper has proposed an incarnation of the coordination language Bach in
% Scala, in the form of an internal Domain Specific Language, named
% B2Scala. It has also proposed an Hennessy-Milner like logic that
% allows for constraining executions. The Needham-Schroeder protocol has
% been modeled with our proposal to illustrate its interest in
% practice.

In the aim of formally verifying security protocols, this
paper has proposed an embedding of the coordination language Bach in
Scala, in the form of an internal Domain Specific Language, named
B2Scala. It has also proposed a logic that
allows for constraining executions. The Needham-Schroeder protocol has
been modeled with our proposal to illustrate its interest in
practice.

The choice for an internal Domain Specific Language has been motivated
by the possibility of taking profit from the Scala eco-system, notably
its type system, while benefiting from all the abstractions offered by
the Bach coordination language. We hope to have convinced the reader
of these two features through the coding of the Needham-Schroeder
protocol. Indeed, on the one hand, the $BSC\_Agents$ coding Alice, Bob
and Mallory mimick the procedures that would have been written
directly in Bach. Moreover the sequential composition operator, the
parallel composition operator and the non-deterministic choice
operator have been used as one would have used them in Bach. This
feature allows to embed the Bach control flow operators in B2Scala.
It is here also worth observing that a similar description could have
been written in a pure process algebra setting like the one used in
the workbenches Scan and Anemone. However type checking is not
supported by these workbenches but is given for free in
B2Scala. Moreover, auxiliary concepts like $public\_key(Y)$ would have
been rewritten as mapping functions, with care for completeness of the
code left to the programmer while it is provided for free in B2Scala
(through completeness verification done by Scala for the match
operation).

On the other hand, the code to be written is a real Scala
code. Examples of that are the definitions of tokens or si-terms,
which are Scala case classes. In that respect, it is worth stressing
that arguments of si-terms need to be declared with a type, which is
verified at compilation time. Moreover, they can be obtained as the
result of a Scala function, as exemplified by the use of
$public\_key(Y)$ in the coding of Alice and Bob (see
Figures~\ref{Alice-in-bscala} and \ref{Bob-in-bscala}).  It is also to
be noted that the $GSum$ construct offers a form of local variable,
binding the Scala and Bach worlds. Take for instance the
first GSum of Figure~\ref{Alice-in-bscala} :

\begin{lstlisting}
GSum( List(bob,mallory), Y => {
     tell(a_running(Y)) * ...
\end{lstlisting}

\noindent
There $Y$ plays the role of a local variable which has to be bound to
$bob$ or $mallory$. Once the value has been decided (by the
\texttt{run\_one} function through the alternative selected for the
choice, see Section~\ref{run-one-ref}), it can be used later in the
code.  Similarly, the second $GSum$ construct allows to bind $WNonce$
to the value selected by the $get$ primitive:

\begin{lstlisting}
GSum( List(na,nb,nm),  WNonce => {
     get( message(Y, alice, encrypt_ii(na,WNonce,pka)) ) * ...
\end{lstlisting}

This being said, our main goal in this paper is to offer a modelling
language to describe and reason on systems, such as the
Needham-Schroeder protocol, rather than a programming language to code
the implementation of the protocol.  In these lines, it is worth
observing that a direct modelling for analysis purposes would not have
been possible in (pure) Scala since we would lack the abstraction
offered by process algebras like Bach.

As reported in~\cite{Mariani-2020}, many coordination languages 
have been implemented, in some cases as stand alone languages, like
Tucson \cite{Tucson}, but mostly as API's of conventional languages,
accessing tuple spaces through dedicated functions or methods, as in
pSpaces \cite{Loreti-et-al}. To the best or our knowledge, B2Scala is
the first implementation of a coordination language as a Domain
Specific Language. We are also not aware of an implementation done in
Scala. However, our work is linked to the work on Caos
\cite{Caos}, which provides, by using Scala, a generic tool to
implement structured operational semantics and to generate intuitive
and interactive websites. In practice, one has however to define the
semantics of the language under consideration by using Scala. In
contrast, we take an opposite approach which already offers an
implementation of the Bach constructs and in which programmers need to
code Bach-like programs in a Scala manner. Moreover we propose a logic
to constraint executions, which is not proposed in \cite{Caos}.

Scafi \cite{Scafi} is another research effort to integrate a
coordination language in Scala. It targets a different line of
research in the coordination community by being focussed on aggregate
computing. Moreover, to the best of our knowledge, no support for
constrained executions is proposed.

This work is a continuation of previous work on the Scan and Anemone
workbenches~\cite{Scan,Anemone}. It differs by the fact that both Scan
and Anemone interpret directly Bach programs. Moreover the PLTL logic
they use is different from the logic proposed in this paper.

As regards the Needham-Schroeder protocol, to our best knowledge, it
has been never been modeled in a coordination language, most probably
because the Coordination community and the one on security are quite
different. Nevertheless it has been modeled in more classical process
algebras. In \cite{Lowe} the author uses CSP and its associated FDR
tool to produce an attack on the protocol. This analysis has been
complemented in \cite{Groote-NDS} by using the mCRL process algebra
and its associated model checker. Our work differs by using a process
algebra of a different nature. Indeed the Bach coordination language
rests on asynchronous communication which happens by information
being available or not on a shared
space. This allows to naturally model messages being put on the
network as si-terms told on the store. Similarly the action of an
intruder is very intuitively modeled by getting si-terms. In contrast,
\cite{Groote-NDS} and \cite{Lowe} use synchronous communication which
does not naturally introduce the network as a communication medium and
which technically forces them to model the intruder by duplicating
Alice and Bob's send and receive actions by intercept and fake
messages.

Our work open several paths for future research. First the synergy
with Scala given by B2Scala offers a natural way of making interfaces
much more user friendly than the one of
Figure~\ref{computation-result}. Second we have only investigated the
use of B2Scala to analyze the Needham-Schroeder protocol. Our current
research aims at exploring the security of other protocols, such as
the Quic protocol. Finally, our logic is used to restrict computations
at run-time without lookahead strategies, which could lead to select
computations that fail later to meet the remaining logic formulae. As
a solution to that problem, we are investigating how statistical model
checking can be married with B2Scala.

% Acknowledgment
\section{Acknowledgment}

The authors warmly thank the anonymous reviewers for their insightful
comments, which greatly contributed to the improvement of this
article. They also thank the University of Namur for its support as
well as the Walloon Region for partial support through the Ariac
project (convention 210235) and the CyberExcellence project
(convention 2110186).

% -----------------------------------------------------------%

\bibliographystyle{eptcs}
\bibliography{biblio}

\end{document}